
\message
{JNL.TEX version 0.9 as of 3/27/86.  Report bugs and problems to Doug Eardley.}



\font\twelverm=cmr10 scaled 1200    \font\twelvei=cmmi10 scaled 1200
\font\twelvesy=cmsy10 scaled 1200   \font\twelveex=cmex10 scaled 1200
\font\twelvebf=cmbx10 scaled 1200   \font\twelvesl=cmsl10 scaled 1200
\font\twelvett=cmtt10 scaled 1200   \font\twelveit=cmti10 scaled 1200

\skewchar\twelvei='177   \skewchar\twelvesy='60


\def\twelvepoint{\normalbaselineskip=12.4pt plus 0.1pt minus 0.1pt
  \abovedisplayskip 12.4pt plus 3pt minus 9pt
  \belowdisplayskip 12.4pt plus 3pt minus 9pt
  \abovedisplayshortskip 0pt plus 3pt
  \belowdisplayshortskip 7.2pt plus 3pt minus 4pt
  \smallskipamount=3.6pt plus1.2pt minus1.2pt
  \medskipamount=7.2pt plus2.4pt minus2.4pt
  \bigskipamount=14.4pt plus4.8pt minus4.8pt
  \def\rm{\fam0\twelverm}          \def\it{\fam\itfam\twelveit}%
  \def\sl{\fam\slfam\twelvesl}     \def\bf{\fam\bffam\twelvebf}%
  \def\mit{\fam 1}                 \def\cal{\fam 2}%
  \def\tt{\twelvett}
  \textfont0=\twelverm   \scriptfont0=\tenrm   \scriptscriptfont0=\sevenrm
  \textfont1=\twelvei    \scriptfont1=\teni    \scriptscriptfont1=\seveni
  \textfont2=\twelvesy   \scriptfont2=\tensy   \scriptscriptfont2=\sevensy
  \textfont3=\twelveex   \scriptfont3=\twelveex  \scriptscriptfont3=\twelveex
  \textfont\itfam=\twelveit
  \textfont\slfam=\twelvesl
  \textfont\bffam=\twelvebf \scriptfont\bffam=\tenbf
  \scriptscriptfont\bffam=\sevenbf
  \normalbaselines\rm}



\def\beginlinemode{\endmode
  \begingroup\parskip=0pt \obeylines\def\\{\par}\def\endmode{\par\endgroup}}
\def\beginparmode{\endmode
  \begingroup \def\endmode{\par\endgroup}}
\let\endmode=\par
{\obeylines\gdef\
{}}
\def\singlespace{\baselineskip=\normalbaselineskip}

\def\oneandahalfspace{\baselineskip=\normalbaselineskip
  \multiply\baselineskip by 3 \divide\baselineskip by 2}
\def\doublespace{\baselineskip=\normalbaselineskip \multiply\baselineskip by 2}

\newcount\firstpageno
\firstpageno=2
\footline={\ifnum\pageno<\firstpageno{\hfil}\else{\hfil\twelverm\folio\hfil}\fi}
\def\toppageno{\global\footline={\hfil}\global\headline
  ={\ifnum\pageno<\firstpageno{\hfil}\else{\hfil\twelverm\folio\hfil}\fi}}
\let\rawfootnote=\footnote		
\def\footnote#1#2{{\rm\singlespace\parindent=0pt\parskip=0pt
  \rawfootnote{#1}{#2\hfill\vrule height 0pt depth 6pt width 0pt}}}
\def\raggedcenter{\leftskip=4em plus 12em \rightskip=\leftskip
  \parindent=0pt \parfillskip=0pt \spaceskip=.3333em \xspaceskip=.5em
  \pretolerance=9999 \tolerance=9999
  \hyphenpenalty=9999 \exhyphenpenalty=9999 }
\def\dateline{\rightline{\ifcase\month\or
  January\or February\or March\or April\or May\or June\or
  July\or August\or September\or October\or November\or December\fi
  \space\number\year}}
\def\received{\vskip 3pt plus 0.2fill
 \centerline{\sl (Received\space\ifcase\month\or
  January\or February\or March\or April\or May\or June\or
  July\or August\or September\or October\or November\or December\fi
  \qquad, \number\year)}}


\hsize=6.5truein
\vsize=8.9truein
\parskip=\medskipamount
\def\\{\cr}
\twelvepoint		
\doublespace		
\overfullrule=0pt	




\def\title			
  {\null\vskip 3pt plus 0.2fill
   \beginlinemode \doublespace \raggedcenter \bf}

\def\author			
  {\vskip 3pt plus 0.2fill \beginlinemode
   \singlespace \raggedcenter}

\def\affil			
  {\vskip 3pt plus 0.1fill \beginlinemode
   \oneandahalfspace \raggedcenter \sl}

\def\abstract			
  {\vskip 3pt plus 0.3fill \beginparmode
   \oneandahalfspace ABSTRACT: }

\def\endtopmatter		
  {\endpage			
   \body}

\def\body			
  {\beginparmode}		

\def\head#1{			
  \goodbreak\vskip 0.5truein	
  {\immediate\write16{#1}
   \raggedcenter \uppercase{#1}\par}
   \nobreak\vskip 0.25truein\nobreak}

\def\beneathrel#1\under#2{\mathrel{\mathop{#2}\limits_{#1}}}

\def\refto#1{$^{#1}$}		

\def\references			
  {\head{References}		
   \beginparmode
   \frenchspacing \parindent=0pt \leftskip=1truecm
   \parskip=8pt plus 3pt \everypar{\hangindent=\parindent}}

\gdef\refis#1{\item{#1.\ }}			

\gdef\journal#1, #2, #3, 1#4#5#6{		
    {\sl #1~}{\bf #2}, #3 (1#4#5#6)}		

\def\prb{\journal Phys. Rev. B, }

\def\prl{\journal Phys. Rev. Lett., }

\def\endreferences{\body}

\def\figurecaptions		
  {\endpage
   \beginparmode
   \head{Figure Captions}
}

\def\endpage			
  {\vfill\eject}

\def\endpaper			
  {\endmode\vfill\supereject}

\def\endit
  {\endpaper\end}


\def\heading				
  {\vskip 0.5truein plus 0.1truein	
   \beginparmode \def\\{\par} \parskip=0pt \singlespace \raggedcenter}

\def\subheading				
  {\vskip 0.25truein plus 0.1truein	
   \beginlinemode \singlespace \parskip=0pt \def\\{\par}\raggedcenter}

\def\tag#1$${\eqno(#1)$$}

\def\align#1$${\eqalign{#1}$$}

\def\aligntag#1$${\gdef\tag##1\\{&(##1)\cr}\eqalignno{#1\\}$$
  \gdef\tag##1$${\eqno(##1)$$}}

\def\overset#1\to#2{{\mathop{#2}^{#1}}}
\def\underset#1\to#2{{\mathop{#2}_{#1}}}


\def\ref#1{Ref.~#1}			
\def\Ref#1{Ref.~#1}			
\def\[#1]{[\cite{#1}]}
\def\cite#1{{#1}}
\def\(#1){(\call{#1})}
\def\call#1{{#1}}
\def\taghead#1{}
\def\frac#1#2{{#1 \over #2}}

\def\12{{1\over2}}

\def\sla{\raise.15ex\hbox{$/$}\kern-.57em}
\def\leaderfill{\leaders\hbox to 1em{\hss.\hss}\hfill}
\def\twiddle{\lower.9ex\rlap{$\kern-.1em\scriptstyle\sim$}}
\def\bigtwiddle{\lower1.ex\rlap{$\sim$}}
\def\gtwid{\mathrel{\raise.3ex\hbox{$>$\kern-.75em\lower1ex\hbox{$\sim$}}}}
\def\ltwid{\mathrel{\raise.3ex\hbox{$<$\kern-.75em\lower1ex\hbox{$\sim$}}}}
\def\square{\kern1pt\vbox{\hrule height 1.2pt\hbox{\vrule width 1.2pt\hskip 3pt
   \vbox{\vskip 6pt}\hskip 3pt\vrule width 0.6pt}\hrule height 0.6pt}\kern1pt}
\def\tdot#1{\mathord{\mathop{#1}\limits^{\kern2pt\ldots}}}

\def\pmb#1{\setbox0=\hbox{#1}%
  \kern-.025em\copy0\kern-\wd0
  \kern  .05em\copy0\kern-\wd0
  \kern-.025em\raise.0433em\box0 }

\def\3he{{$^3${\rm He}}}

\def\slD{\raise.15ex\hbox{$/$}\kern-.57em\hbox{$D$}}
\def\dsl{\raise.15ex\hbox{$/$}\kern-.57em\hbox{$\Delta$}}
\def\slp{{\raise.15ex\hbox{$/$}\kern-.57em\hbox{$\partial$}}}
\def\nsl{\raise.15ex\hbox{$/$}\kern-.57em\hbox{$\nabla$}}
\def\sla{\raise.15ex\hbox{$/$}\kern-.57em\hbox{$\rightarrow$}}
\def\slla{\raise.15ex\hbox{$/$}\kern-.57em\hbox{$\lambda$}}
\def\slb{\raise.15ex\hbox{$/$}\kern-.57em\hbox{$b$}}
\def\lnp{\raise.15ex\hbox{$/$}\kern-.57em\hbox{$p$}}
\def\lnk{\raise.15ex\hbox{$/$}\kern-.57em\hbox{$k$}}
\def\lnK{\raise.15ex\hbox{$/$}\kern-.57em\hbox{$K$}}
\def\lnq{\raise.15ex\hbox{$/$}\kern-.57em\hbox{$q$}}
\def\lnA{\raise.15ex\hbox{$/$}\kern-.57em\hbox{$A$}}
\def\lna{\raise.15ex\hbox{$/$}\kern-.57em\hbox{$a$}}
\def\lnB{\raise.15ex\hbox{$/$}\kern-.57em\hbox{$B$}}

\def\cL{{\cal L}}


\def\pmb#1{\setbox0=\hbox{$#1$}%
\kern-.025em\copy0\kern-\wd0
\kern.05em\copy0\kern-\wd0
\kern-.025em\raise.0433em\box0 }

\def\q2{{Q^2}}
\def\gtwid{\raise.3ex\hbox{$>$\kern-.75em\lower1ex\hbox{$\sim$}}}
\def\ltwid{\raise.3ex\hbox{$<$\kern-.75em\lower1ex\hbox{$\sim$}}}
\def\12{{1\over2}}
\def\part{\partial}

\def\low#1{\lower.5ex\hbox{${}_#1$}}

\def\psl{\raise.15ex\hbox{$/$}\kern-.57em\hbox{$\partial$}}
\def\partt{\raise.15ex\hbox{$\widetilde$}{\kern-.37em\hbox{$\partial$}}}
\def\r#1{{\frac1{#1}}}

\def\topppageno1{\global\footline={\hfil}\global\headline
={\ifnum\pageno<\firstpageno{\hfil}\else{\hss\twelverm --\ \folio
\ --\hss}\fi}}

\def\toppageno2{\global\footline={\hfil}\global\headline
={\ifnum\pageno<\firstpageno{\hfil}\else{\rightline{\hfill\hfill
\twelverm \ \folio
\ \hss}}\fi}}

\catcode`@=11
\newcount\r@fcount \r@fcount=0
\newcount\r@fcurr
\immediate\newwrite\reffile
\newif\ifr@ffile\r@ffilefalse
\def\w@rnwrite#1{\ifr@ffile\immediate\write\reffile{#1}\fi\message{#1}}

\def\writer@f#1>>{}
\def\referencefile{
  \r@ffiletrue\immediate\openout\reffile=\jobname.ref%
  \def\writer@f##1>>{\ifr@ffile\immediate\write\reffile%
    {\noexpand\refis{##1} = \csname r@fnum##1\endcsname = %
     \expandafter\expandafter\expandafter\strip@t\expandafter%
     \meaning\csname r@ftext\csname r@fnum##1\endcsname\endcsname}\fi}%
  \def\strip@t##1>>{}}

\def\citeall#1{\xdef#1##1{#1{\noexpand\cite{##1}}}}
\def\cite#1{\each@rg\citer@nge{#1}}     

\def\each@rg#1#2{{\let\thecsname=#1\expandafter\first@rg#2,\end,}}
\def\first@rg#1,{\thecsname{#1}\apply@rg}       
\def\apply@rg#1,{\ifx\end#1\let\next=\relax
\else,\thecsname{#1}\let\next=\apply@rg\fi\next}

\def\citer@nge#1{\citedor@nge#1-\end-}  
\def\citer@ngeat#1\end-{#1}
\def\citedor@nge#1-#2-{\ifx\end#2\r@featspace#1 
  \else\citel@@p{#1}{#2}\citer@ngeat\fi}        
\def\citel@@p#1#2{\ifnum#1>#2{\errmessage{Reference range #1-#2\space is bad.}%
    \errhelp{If you cite a series of references by the notation M-N, then M and
    N must be integers, and N must be greater than or equal to M.}}\else%
 {\count0=#1\count1=#2\advance\count1
by1\relax\expandafter\r@fcite\the\count0,%

  \loop\advance\count0 by1\relax
    \ifnum\count0<\count1,\expandafter\r@fcite\the\count0,%
  \repeat}\fi}

\def\r@featspace#1#2 {\r@fcite#1#2,}    
\def\r@fcite#1,{\ifuncit@d{#1}
    \newr@f{#1}%
    \expandafter\gdef\csname r@ftext\number\r@fcount\endcsname%
                     {\message{Reference #1 to be supplied.}%
                      \writer@f#1>>#1 to be supplied.\par}%
 \fi%
 \csname r@fnum#1\endcsname}
\def\ifuncit@d#1{\expandafter\ifx\csname r@fnum#1\endcsname\relax}%
\def\newr@f#1{\global\advance\r@fcount by1%
    \expandafter\xdef\csname r@fnum#1\endcsname{\number\r@fcount}}

\let\r@fis=\refis                       
\def\refis#1#2#3\par{\ifuncit@d{#1}
   \newr@f{#1}%
   \w@rnwrite{Reference #1=\number\r@fcount\space is not cited up to now.}\fi%
  \expandafter\gdef\csname r@ftext\csname r@fnum#1\endcsname\endcsname%
  {\writer@f#1>>#2#3\par}}

\def\ignoreuncited{
   \def\refis##1##2##3\par{\ifuncit@d{##1}%
     \else\expandafter\gdef\csname r@ftext\csname
r@fnum##1\endcsname\endcsname%

     {\writer@f##1>>##2##3\par}\fi}}

\def\r@ferr{\endreferences\errmessage{I was expecting to see
\noexpand\endreferences before now;  I have inserted it here.}}
\let\r@ferences=\references
\def\references{\r@ferences\def\endmode{\r@ferr\par\endgroup}}

\let\endr@ferences=\endreferences
\def\endreferences{\r@fcurr=0
  {\loop\ifnum\r@fcurr<\r@fcount
    \advance\r@fcurr by 1\relax\expandafter\r@fis\expandafter{\number\r@fcurr}%
    \csname r@ftext\number\r@fcurr\endcsname%
  \repeat}\gdef\r@ferr{}\endr@ferences}


\let\r@fend=\endpaper\gdef\endpaper{\ifr@ffile
\immediate\write16{Cross References written on []\jobname.REF.}\fi\r@fend}

\catcode`@=12

\citeall\refto          
\citeall\ref            %
\citeall\Ref            %



\def \q {\Psi^{(2)}([z^+_i],[z^-_i])}
\def \r {\Psi_m^{(1)}([z^+_i],[z^-_i])}
\def \g {\Psi_m([z^+_i],[z^-_i])}
\def \x {\xi_{\sigma}}

\def \zu {z^+_i}
\def \zd {z^-_i}

\title Spin Singlet Quantum Hall Effect and Nonabelian Landau-Ginzburg Theory
\author Alexander Balatsky
\affil Los Alamos National Laboratory
Theoretical Division, MS B262
Los Alamos, NM 87544
\centerline{and}
\affil Landau Institute for Theoretical Physics, Moscow, USSR.

\abstract

In this paper we present a theory of  Singlet Quantum
Hall Effect (SQHE). We show that the Halperin-Haldane SQHE wave function can
be written in the form of a product of a wave function for charged semions in
a magnetic
field and a wave function for the Chiral Spin Liquid of neutral spin-$\12$
semions.
We introduce  field-theoretic model in which the electron operators
are factorized in terms of charged spinless semions (holons) and
neutral spin-$\12$ semions (spinons). Broken time reversal symmetry
and short ranged spin
correlations lead to $SU(2)_{k=1}$ Chern-Simons term in Landau-Ginzburg
 action for SQHE phase.
We construct appropriate coherent states for SQHE phase and show the existence
 of $SU(2)$ valued
gauge potential. This potential appears as a result of ``spin rigidity" of
 the ground state
against any displacements of nodes of wave function from positions
of the particles and reflects
the nontrivial monodromy in the presence of these displacements.
We argue that topological structure of
$SU(2)_{k=1}$ Chern-Simons theory unambiguously dictates {\it semion}
statistics of spinons.

 \body
\vskip .15in
\noindent PACS No.73.20.Dx; 11.15.-q; 75.10.Jm; 74.65.+n
\endtopmatter

\endpage
\head{Content}

\noindent I. INTRODUCTION.

\noindent i) General remarks.

\noindent ii) Spin polarized Fractional QHE and Landau-Ginzburg Theory.

\noindent II. SPIN SINGLET QUANTUM HALL EFFECT and LANDAU-GINZBURG THEORY.

\noindent i) Halperin-Haldane Wave  Function of SQHE  and Slave Semion
Decomposition.

\noindent ii) Coherent States for Singlet QHE.

\noindent iii) $SU(2)_{k =  1}$ Chern-Simons Theory as a LG Functional for
Singlet
QHE.

\noindent III. CONCLUSION.

\endpage

\head{ I.Introduction.}
\vskip 1pc

\noindent
{\bf i) General remarks}
\vskip 0.5pc

It has been assumed from the beginning of the theory of Fractional Quantum Hall
Effect (FQHE), that the magnetic field, which has to be strong enough to
produce the
relevant Landau quantization, leads to large Zeeman splitting. Large body of
physical theories of FQHE assumed spins of electrons to be polarized completely
(which is equivalent to consideration of spinless electrons in the lowest
Landau
level).

It has been pointed out first by Halperin\refto{Halperin} that this is not
always the case.
Zeeman splitting is given by E$_{Zeeman} = g\cdot\mu_B\cdot H$, and Larmour
energy is
E$_{Larmour} = eH/\hbar c$. The ratio of these two energies depends on the
factor E$_{Zeeman}/E_{Larmour} = g\cdot {m^\ast\over m_o}$, where m$^\ast$ is
the effective mass of electron, and g - is the g-factor. The ratio of
$m^\ast/m_o$ in the Si/SiO$_2$ structures is quite small $m^\ast/m_o \simeq$
0.07, and g can be as low as 1/4.

We find, thus, that at least in low enough magnetic fields B $\sim$ 1 T, for
some materials the ratio
${E_{Zeeman}\over E_{Larmour}} \simeq$ 0.017 is quite small. Thus it is a good
approximation in this
case to neglect Zeeman splitting and consider all states in the Hilbert space
of the problem as
doubly degenerate due to spin.

Within these assumptions one has to consider the spin unpolarized QHE phase. We
will consider below
the case of spin singlet QHE phase (SQHE).

Experimentally there is  evidence that spin singlet QHE phases are present at
some filling
factors, see for example.\refto{5/2}

In this article we will consider the Landau-Ginzburg theory of singlet QHE and
how it is connected  with nonabelian, namely SU(2)$_{k=1}$ for spin S=1/2,
Chern Simons
theory as a natural generalization of the Chern Simons theory for spin
polarized
case. We will show how the SU(2) valued gauge potential naturally appears in
the
context of spin coherent states for SQHE \refto{BF}.

But before considering spin unpolarized case we will summarize briefly the most
important features of
the Landau-Ginzburg theory for spin polarized case.
\vskip 1pc

\noindent
{\bf ii) Spin polarized FQHE and Landau Ginzburg theory}
\vskip 0.5pc

Soon after experimental discovery of the Fractional QHE (FQHE)$^2$ Laughlin
proposed variational wave
function which describes the incompressible electron liquid in 2D in external
magnetic field at
fractional filling factors, which naturally leads to the ``fractional
statistics'' of the
quasiparticles.\refto{Laughlin}. The {\it holomorphic} structure of the
Laughlin state relies
essentially on the fact that the coordinate space of electron liquid is 2D.

$$
\Psi_L (r_i)~=~\prod_{i<j}~(z_i - z_j)^m~{\rm exp}\left(-~{1\over
4}~|z_i|^2\right) \eqno(I.1) $$

The conductivity tensor of this state is

$$
\sigma_{xy}~=~{e^2\over h}~{1\over m} \eqno(I.2)
$$

\noindent
where m- is an odd integer. The last observation proved to be crucial for the
construction of the
phenomenology of the fractional Quantum Hall Effect (FQHE).

Namely it has been noticed by Girvin and MacDonald ,\refto{GM} and
subsequently by others\refto {ZHK,R} that broken time reversal invariance and
parity leads to the possibility for  parity noninvariant terms in the
Landau-Ginzburg (LG) functional for electrons in FQHE phase. Physics of the
Laughlin state Eq. (I.1) is dictated by the fact that in this correlated state
each electron at point Z$_i$ is confined with m quanta of magnetic flux
$\varphi_o = {hc\over e}$. Effect of the Chern-Simons term in the LG functional
of FQHE is to reinforce the constraint, that {\it density of particles is
proportional to the local flux value of some gauge field}, which we will call
the
statistical gauge field , A$_\mu$:

$$
\varphi^\ast\varphi - \langle\varphi^\ast\varphi\rangle~=~{k\over
2\pi}~\epsilon_{ij}~F^{ij}
\eqno(I.3) $$
With $\varphi$-scalar classical field associated with the order parameter in
FQHE.\refto
{R} The reason for introducing extra statistical gauge field is  to take into
account density
fluctuations and associated fluctuations of the phase of the wave function. At
the same time the
external magnetic field is essentially constant.

The LG functional has the form:\refto{GM,ZHK,R}

$$
\eqalignno{{\cal L}~=~&\int d^2 x dt~\varphi^\ast (i\partial_o - e{\cal
A}_o - A_0)\varphi~-~{1\over 2m} |(i\partial_i - e{\cal A}_i - A_i)\varphi
|^2\cr
\noalign{\vskip 0.5pc}
&+~V(\varphi)~+~{k\over 4\pi}~A_\mu~\partial_v~A_\lambda~\epsilon^{\mu
v\lambda}&(I.4)\cr} $$
with $k = {2\pi e^2\over h} {1\over m}, {\cal A}_\mu$ is electromagnetic
potential, and
$V(\varphi) = -\alpha\varphi^2 + \beta\varphi^4$ is the potential which fixes
the amplitude
of the order parameter $\varphi$. The last term in Lagrangian ${\cal L}$ is a
Chern Simons
terms for the statistical gauge potential $A_\mu$ with the group U(1).
Variation of ${\cal
L}$ over electromagnetic potential leads to the expression for the transverse
conductivity given by Eq. (I.2). More detailed analysis of the LG theory for
FQHE in the
spin polarized case can be found in.\refto{GM,ZHK,R,QHE}

Examining LG theory of spin polarized case we can make
two general statements

a) The holomorphic structure of the wave function and closely related to its
presence of
strong magnetic field in the system allows us to write parity and time reversal
noninvariant terms, such as Chern Simons for some statistical gauge field.

b) This gauge field obeys the constraint that the  {\it density is proportional
to the flux
of this gauge field}, like Eq. (I.3). We will argue below that these two
statements are also true for the case of LG theory of SQHE. The only essential
difference comes from the fact that the gauge group, corresponding to
statistical gauge field, will be SU(2). Instead of density operator playing
role
of the generator of the flux, the spin will be the generator of the spin gauge
field flux. In contrast to the U(1) group SU(2) Chern-Simons theory is true
topological  theory. As a result of quantization of the coefficient in the
Chern-Simons term, we will find that the only fractional statistics of
excitations for SU(2) at level k=1 Chern Simons theory will be {\it semion}
statistics.

\endpage

\head{II. Spin
Singlet Quantum Hall Effect and Landau-Ginzburg Theory.}

{\bf i) Halperin-Haldane Wave  Function of SQHE  and Slave Semion
Decomposition.}

In this paragraph we consider the physical properties of the singlet Quantum
Hall Effect states, given by the Halperin-Haldane wave function
\refto{Halperin},\refto{Hal}:
$$\eqalign{ \g = \prod_{i<j}(z^+_i - z^+_j)^{m+1} (z^-_i -
z^-_j)^{m+1} (z^+_i  -  z^-_j)^{m} \cr e^{-{1\over{4}}\sum_i|z^+_i|^2
-{1\over{4}}\sum_i|z^-_i|^2} \cr } .\eqno(II.1)$$
where the set of coordinates $z^+_i, i = 1, \dots ,N$ corresponds to the spin
$\uparrow$ electrons, and $z^-_i, i = 1, \dots ,N$ corresponds to
the spin $\downarrow$
electrons and $m$ is an even integer. In this case $\g$ satisfies the Fock
cyclicity condition. In this state, the eigenvalue of the total spin operator
is
$S = 0$ and the $z$-component of the spin also has eigenvalue $S_z = 0$.
This kind of  wave functions naturally appears in the
consideration of the spin unpolarized states in the Quantum Hall Effect (QHE)
phase.

In contrast to the spin polarized states, in this case we need to describe the
charge sector of the SQHE phase as well as the spin sector. By inspecting the
structure of this wave function one finds that it has the simple but very
important property that the spin and charge degrees of freedom are
factorized. The total wave function $\g$ can be written as a product of the
charge wave function $\r $ and spin wave function $\q $.
Below we will discuss the properties of the charge and spin wave functions
separately. At the end we will put them together again by imposing
the constraint that the positions of the charges coincides with those of
the spins. This property is strongly reminiscent of the charge and spin
separation present in models of Strongly Correlated Electron systems in the
context of theories of high temperature superconductors\refto{Str.Cor}.

The wave function is factorized in the following manner \refto{BF}:
$$\g = \q \r .\eqno(II.2)$$
with
$$\eqalign{ \r   = \prod_{i<j}(z^+_i - z^+_j)^{m+1/2} (z^-_i
- z^-_j)^{m+1/2} (z^+_i  - z^-_i)^{m+1/2} \cr  e^{-{1\over{4}}
\sum_i|z^+_i|^2 -{1\over{4}}\sum_i|z^-_i|^2} \cr }  . \eqno(II.3)$$
$$\q =  \prod_{i<j}(z^+_i - z^+_j)^{1/2} (z^-_i -
z^-_j)^{1/2} (z^+_i  - z^-_i)^{-1/2}  .\eqno(II.4)$$
Why does this decomposition make sense?. The plasma
analogy, when applied to $\r $, shows that this state is described by a one
component plasma, in which the particles at points $z^+_i$ and
$z^-_i$ have equal charge:
$$ \eqalign {|\r|^2 =   \exp((2m+1)(\sum_{i<j} \ln|z^+_i - z^+_j| + \sum_{i<j}
\ln|z^-_i - z^-_j| + \sum_{i,j} \ln|z^+_i - z^-_j|) \cr
-1/2\sum_i|z^+_i|^2 - 1/2\sum_{i}|z^-_i|^2 ) \cr }  . \eqno(II.5)$$
We regard $\r $ as the wave function for the charge degrees of freedom.

If we apply the same plasma analogy to the wave function $\q$ we get \refto{G}:
$$ |\q|^2 = \exp(\sum_{i<j} \ln|z^+_i - z^+_j| + \sum_{i<j}
\ln|z^-_i - z^-_j| - \sum_{i,j} \ln|z^+_i -
z^-_ j|)
        .\eqno(II.6)$$
and we can easily see that $\q$ corresponds to a two-component plasma, where
the effective charge of the particles $q$ is given by the spin projection $q =
2
s_z = \pm1$. It is natural to consider $\q$ as the wave function of the spin
degrees of freedom.

We will show below that $\r $ can be regarded as a wave function for  semions
in an external external magnetic field. From Eq.(II.3) we conclude that, for
any
$m$, $\r $ describes
particles with semion statistics: any exchange of two of them leads to a
change of phase of $ \pi (m + 1/2)$ and, if $m$ is even, this particles are
semions. From
the same considerations it follows that $\q$ represents a two-component semion
gas. The sign of the spin projection $s_z$ determines  the effective
phase change in any interchange of two particles $q_1q_2 \pi/2$, where
$q_1,q_2$ are  $\pm 1$ for spin $\uparrow,\downarrow$. This model with two
component semions was considered in \refto{BK,G}. In particular,
Girvin et al. \refto{G} have pointed out that the state described by
the wave function $\q$ is a local spin singlet due to the plasma screening of
any charge.

The decomposition of Eq.(II.2) can be represented in terms of the slave
{\it semion} operators:
$$\psi_{\sigma}(r) = \varphi(r) \x(r) .\eqno(II.7)$$
where $\psi_{\sigma}(r)$ is the electron operator, $\varphi(r)$ is a charge $e$
spinless  semion operator, $\x$ is a spin 1/2 charge-neutral semion operator,
$\sigma $ is a spin index, and we assume that $[\varphi(r),\xi(r)] = 0$.

In principle this decomposition is neither better nor worse than any other
slave boson or slave fermion factorization, like the ones that
are commonly used in theories of
strongly correlated systems. The choice of any particular
decomposition of the initial electron operator is purely a matter of
convenience. Our choice is
motivated by the simplicity of the physical picture that we get in the end.

In Mott-Hubbard insulators, the strong correlations force the
constraint of single particle occupancy. In the case of the SQHE, the origin of
the strong correlations is the drastic reduction of phase space due to the
presence of a strong magnetic field: the kinetic energy is quenched and the
interactions dominate. In close analogy with the Mott-Hubbard problem, we argue
that in the Singlet Quantum Hall Effect the spin and charge degrees of freedom
are separated in the sense of the decomposition of Eq.(II.7). Here too, a gauge
symmetry arises as a result of this factorization. This gauge symmetry means
that the {\it relative} phase between charge and spin states
is not a physically observable degree of freedom. The SQHE wave function is a
singlet under this gauge symmetry. However, the decomposition Eq.(II.7)
requires
that the entire spectrum of states must be singlets under this gauge symmetry.
Given the close analogy with the Mott-Hubbard problem, we will refer to this
symmetry as the RVB gauge symmetry. The presence of this RVB gauge symmetry
gives rise to an RVB gauge field which puts the charge and spin
semions together to form the allowed physical states. Thus, although the
wave functions of all the states can be factorized as a product
of a  {\it charge} and {\it spin} wave functions,
 there is no separtaion of spin and charge in this system. In
consequence,
the system has a gap to {\it all} excitations and it is {\it incompressible}.
The factorized form of the SQHE
wave function, Eq.(II.2), appears to suggest that there may be a gapless
neutral spin excitation which would lead to {\it compressibility}.
Because the RVB gauge charge is confined, these excitations
are not a  part of the physical spectrum. It is important to stress that the
incompressibility results entirely from the charge sector.

Perhaps the simplest way to see this is to consider the wave function of
quasiparticle (qp) in the first quantized representation, as it has been done
in \refto{Hal}. for example, for the qp of spin 1/2 with $s_z = -1/2$ at point
$z_0$:
$$ \Psi_{z_0, \downarrow} ([\zu], [\zd]) = \prod_i (z^+_i - z_0) \g
\eqno( II.8)$$
The form of this wave function indicates that the creation of the qp is
equivalent to the creation of the extra zero at point $z_0$ for the wave
function of the particles with the spin $s_z = +1/2$ projection. By using the
plasma analogy it is easy to conclude that this zero is equivalent to the qp of
spin $s_z = - 1/2$ with charge $e = {1\over{2m + 1}}$.

Now we will explicitly show that the wave function of the qp in the SQHE can be
represented as a composite excitation of neutral spinon with $s = 1/2$ and of
the spinless holon with charge $e = {1\over{2m + 1}}$. We can rewrite
$\Psi_{z_0, \downarrow} $ as :
$$\eqalign{ \Psi_{z_0, \downarrow} ([\zu], [\zd]) =&
\prod_{i} (z^+_i - z_0)^{1/2} (\zd - z_0)^{1/2} \r \cr
  &\prod_{i} (z^+_i -
z_0)^{1/2} (\zd - z_0)^{-1/2} \q } \eqno(II.9)$$
The first product $\Psi^{(1)}_{ z_0} = \prod_{i} (\zu - z_o)^{1/2}
(\zd - z_0)^{1/2} \r $ is nothing more then the holon excitation in
the one component plasma, corresponding to the $\r$. From this follows that the
effective charge of the holon is $ e = {1/2\over {m + 1/2}} = {1 \over{2m+1}}$.
The second product $ \Psi^{(2)}_{z_0 \downarrow} = \prod_{i} (\zu -
z_0)^{1/2} (\zd - z_0)^{- 1/2} \q $
is the spinon excitation, corresponding to the extra spin
$s_z =  - 1/2$ excitation, created at point $z_0$.

There is an apparent problem with the identification of the sign of the spin
projection for the excitation $\Psi^{(2)}_{ z_0 \downarrow}$. By the plasma
analogy
the fictitious
spin
1/2 at point $z_0$  has the same projection as the spinons
at points $z_i$, i.e. $s_z = +1/2$. But then, due to the plasma screening in
the two component plasma, the real spinons will screen out this fictitious
spin, thus creating the $s_z = - 1/2$ cloud of real spinons, centered at point
$z_0$. This is precisely the reason why the spin projection of the excitation
$\Psi^{(2)}_{ z_0 \downarrow}$ is down.

Once this confusing point has been clarified, we come to the statement that the
spin 1/2 charge $ e = {1\over{2m + 1}}$ qp can be represented as a product of
the spinon and holon qp created at the same point $z_0$:
$$ \Psi_{z_0 \downarrow} = \Psi^{(1)}_{ z_0} \Psi^{(2)}_{ z_0 \downarrow}
\eqno(II.10)$$
The Eq.(II.10) is a decomposition in Eq.(II.7) written in the first quantized
representation.

We find that the slave semion decomposition (II.7) for the SQHE is valid not
only in the ground state but for the  qp excitations as
well. Clearly the argument given above can be generalized trivially for the
case of $n$ qp. The fact that we need to put our spinon and holon on the same
place explicitly indicates that these excitations with opposite RVB charge are
confined to form an RVB neutral object, only allowed as the physical state
\refto{BF}.

Thus we showed that the slave semion decomposition Eq.(II.7) is quite natural
way to distinguish the physics in the charge and spin sector of SQHE. This
{\it factorization} ( but not {\it separation}) can be observed for any
state in the Hilbert space of SQHE.

\noindent
{\bf ii) Coherent states for SQHE}
\vskip 0.5pc

In this section we will introduce  coherent states for SQHE. Originally
coherent states were
introduced by Read\refto{R}for FQHE
in the spin polarized case.  The generalization of this
construction towards spin unpolarized case is straightforward. In this
construction we will use
analogy with the coherent states for superconductors.

Suppose we have a system, in which some composite operator, involving few
particle operators acquire
the nonzero expectation value. An example of such an operator is
 a superconducting order
parameter

$$
\Delta~=~\langle N|\psi^+_{\uparrow k}~\psi^+_{\downarrow-k}|N+2\rangle
\eqno(II.11)
$$where $\psi^+_{\alpha k}$ is the single particle operator with spin $\alpha$
and momentum $\vec
k, and |N>$ - is the wavefunction of superconductor with N particles.

Clearly, the two particle operator such as in Eq. (II.11) can not have nonzero
expectation value in the
state with the fixed number of particles $|N>, \langle N|\psi^+_{\alpha
k}~\psi^+_{\beta-k}|N\rangle\equiv$ 0, because this object is not gauge
invariant under global U(1)
gauge transformations. In thermodynamic limit we usually consider the system
with fixed chemical
potential and indefinite number of particles which allows the operator to have
a nozero
expectation value. This kind of states allows us to get nonzero expectation
value for the two
particle operator. The phase of the order parameter $\Delta$ has to be well
defined in
superconductor, and taking into account that density operator $\hat n_k =
\psi^+_{\alpha k}
\psi_{\alpha k}$ and phase $\varphi_k$ are canonically conjugated variables
$[\hat n_k,
\varphi_{k\prime}] = i\delta(k-k')$, we find that the superposition of states
with indefinite
number of particles but with fixed phase are natural for considering
superconductors.

These coherent states

$$
|\theta> = \sum^{\infty}_{N=1}~\beta_N~e^{iN\theta}~|N> \eqno(II.12)
$$where $\theta$ is the phase, $\beta_N$ is some weight which is peaked around
macroscopical value
$N=\overline {N}$ with variance $\Delta N \sim \overline {N}^{1/2}$. In this
basis one
easily find that the order parameter becomes a classical field:

$$
\Delta~=~\langle\theta|\psi^+_{k\alpha}~\psi^+_{-k\beta}|\theta\rangle~=~|\Delta_o|e^{i\eta}
\eqno(II.13)
$$with the well defined amplitude $|\Delta_o|$ and phase $\eta$.

 From this transparent example we conclude that if the order parameter as an
operator involves few
particle operators the appropriate basis for consideration of this phase are
the coherent states which
are  coherent superposition of states with different number of particles.

Application of coherent states for construction of the LG theory of polarized
FQHE has been done by N.
Read,\refto{R} see also.\refto{MR} Here we will follow these ideas to construct
the coherent states
for spin unpolarized QHE.

Define states $|N_+, N_->$ as:

$$
|N_{+},N_->~\equiv~\g \eqno(II.14)
$$where $N_\pm$ is the number of particles with up(down) spin. Introduce the
coherent states:

$$
|\theta_+,
\theta_->~=~\sum_{N_\pm}~\beta_{N+}~\beta_{N-}~e^{-iN_+\theta_+-iN_-\theta_-}~|N_+,N_->
\eqno(II.15)
$$where $\beta_{N\pm}$ are some weights with $\langle N_\pm\rangle = \overline
{N}$, and some
variance. $\Delta N_\pm \sim \overline {N}^{1/2}$. This state is with undefined
$S_z$ and undefined
number of particles. The following composite
operator acquires the nonzero expectation value in the $|\theta_+,\theta_->$
state
\refto{MR}:

$$
\land_+(z)~=~\psi^+_+ (z)U^{m+1}_+~(z)~U^m_-~(z) \eqno(II.16a)
$$

$$
\langle\theta_+, \theta_-|~\land_+|~\theta_+,\theta_-\rangle~=~{\rm const}
\eqno(II.16b)
$$
where $U_\pm (z)$ is the flux operator, which produces a node in the wave
function
$|\theta_+, \theta_->$:

$$
U_\pm (z)~=~\prod_i~(z-z^\pm_i) \eqno(II.17)
$$

 And state $|\theta_+, \theta_->$ is simply the condensate of the composite
operators:

$$
|\theta_+, \theta_->~\sim~\sum_{N_\pm}~\beta_{N+}~\beta_{N-} \left(\int
d^2\,z^+\,\land_+
(z^+)\right)^{N_+}~\left(\int d^2z^-~\land_-~(z^-)\right)^{N_-}~ e^{iN_\pm
\theta_\pm}~|0>
\eqno(II.18) $$

Eqs. (II.16)-(II.18) are just the  mathematical expression of the physically
transparent fact that
in the Halperin-Haldane state the electrons of spin up and down are confined
with the zeros of the
wave function. For example each electron of spin up is confined with the (m+1)
st power of zero in the
wave function for all other spin up electrons and m-st power for electrons of
spin down.

As we are mainly concerned with spin dynamics of SQHE, we consider coherent
states and
the appropriate order parameter for the spin wave function $\psi^{(2)}
([z^+_i],[z^-_i])$ in Eq.
(II.4). It has been argued\refto{B} that this wave function describes the spin
1/2 Chiral Spin
Liquid (CSL) state. It follows from this that the spin dynamics of SQHE and
spin dynamics of CSL
phase are closely related. However there is one principal difference between
spin
excitations allowed in SQHE and CSL: the only spin excitations allowed in the
bulk of SQHE
are gaped spin S=1 spin waves, while there are spin 1/2 spinons in the CSL
state. This
difference comes from the fact that spinons in the bulk of SQHE sample are
confined because
of analiticity of the wave function, as we mentioned earlier.

The composite operators which condense in the CSL state are

$$
\land^{CSL}_+(z)~=~\psi^+_+ (z)~U^{1/2}_+ (z)~U^{-1/2}_- (z) \eqno(II.19)
$$

Coherent states natural for CSL are given by Eq. (II.18) with obvious
substitution $\land_\pm \to
\land^{CSL}_\pm$.

The composite operator $\land^{CSL}_\pm$ describes the condensation of the flux
$\pm \pi$ on the
particles with spin $S_z = \pm$. Half flux condensation implies that semion
statistics of excitations
should be expected in this state, and indeed as it is known that spinons are
fractional statistics
excitations in this state.\refto{B}

Using these facts we are now ready to construct the LG theory of SQHE phase.
\vskip 1pc

\noindent
{\bf iii)~SU(2)$_{k=1}$ Chern Simons Theory as a LG Functional for Singlet
QHE.}
\vskip 0.5pc

Below we will consider only spin aspect of the LG theory of SQHE, and thus only
the
neutral excitations will be considered. Because of decomposition Eq. (II.2),
the charge
sector can be treated analogously to the derivation of LG theory for spin
polarized FQHE.

Due to the charge-spin factorization in the Halperin-Haldane state, we will use
composite operator factorization

$$
\land_\pm (z)~=~\land^{CSL}_\pm (z) \cdot \land^{charge} (z) \eqno(II.20)
$$
with obvious form for $\land^{charge} (z)$ which is independent on spin
indexes, and
analogously for coherent states:

$$
|\theta_+, \theta_->~=~|\theta_+, \theta_->_{spin}\cdot
|\theta_+,\theta_->_{charge}
\eqno(II.21)
$$
Obviously the operator, relevant for spin dynamics is $\land^{CSL}_\pm (z)$ and
the
wavefunction containing all information about spin configurations of electrons
is
$|\theta_+,\theta_->_{spin}$, defined as:

$$
\eqalignno{|\theta_+\theta_->_{spin}~&=~\sum_{N_+,N_-}~\beta_{N_+}\beta_{N_-}~e^{iN_\pm
\theta_\pm}\cr \noalign{\vskip 0.5pc}
&\times~\left(\int d^2z_+~\land^{CSL}_+~(z_+)\right)^{N_+} \left(\int d^2
z_-~\land^{CSL}_- (z_-)\right)^{N_-} |0>&(II.22)\cr}
$$
with the same  notations used as  was used in the definition of
$|\theta_+,\theta_->$ Eq. (II.15).
In what follows we will drop the ``CSL'' from the spin composite operator
$\land^{CSL}_\pm (z)$
and ``spin'' from the spin part of the wavefunction
$|\theta_+\theta_->_{spin}$.

Crucial object in deriving the LG functional is the gauge potential, which
appears as a
result of displacement of zero of the wavefunction $\psi^{CSL} ([z^+_i],
[z^-_i])$ from
the position of electron to which this zero is confined. Namely, consider the
following
operator:

$$
\land_\mu (z,z^\prime)~=~\prod_{\mu\prime} \psi^+_\mu (z)~U^{1/2
\mu\cdot\mu\prime}_{\mu\prime} (z') \eqno(II.23)
$$
where $\mu,\mu^\prime = \pm$, and at z = z$'$ this operator is just the
$\land_\mu (z)$,
discussed above.
Below we will assume that operators $\land_\mu (z,z^\prime)$ are normalized by
the factor

 $N_\mu = \langle\theta_+, \theta_-|~\prod_{\mu\prime} U^{1/2
\mu\cdot\mu\prime}_{\mu\prime} (z)|\theta_+,\theta_-\rangle^{-1}$ what will be
taken into
account in the expansion of the composite operator. The displacement z - z$'$
between the point
at which the particle was created and the point at which the zeros of wave
function are located
may lead to nontrivial monodromy properties of the wave function in the
presence of such
displacements. Physically this nontrivial monodromy of particle wave function
around closed
contour C, enclosing such  displacements,  leads to a frustration of the
wavefunction. This in
turn leads to the increase of energy. The system prefers the ground state in
which the
zeros of wavefunction are confined to the positions of the particles.\refto{GM}
The gauge
potential which reflects nontrivial monodromy of probe particle in the
nonhomogeneous case
in spin polarized FQHE, see Eq. (I.3), appears naturally in this analysis .

The difference between polarized FQHE and SQHE is that in SQHE phase pure spin
distortions can produce gauge potential, even if  the charge fluctuations do
not lead
to any U(1) gauge potential as in Eq. (I.3). This concept of binding  zeros of
spin wave
function with the particles  was called ``spin rigidity'' in the case of CSL to
stress the
topological effect caused by displacements between zeros of wave function and
positions of
particles.\refto{B2}

Here we shall see that the same ``spin rigidity'' of the SQHE ground state in
the spin sector
leads to SU(2) valued gauge potential $\hat A_x = i\cdot
A^i_x\cdot\sigma^i_{\alpha\beta},
\sigma^i_{\alpha\beta}$ are Pauli matrices. This gauge potential measures the
nontriviality
of monodromy of spin wave function. Define $\hat A_\pm = \hat A_x \pm i\hat
A_y$ and:

$$
iA^{\nu\mu}_-~=~\lambda~\int~{d^2z^\prime\over z'-z}~\langle\land^+_\mu
(z',z)~\land_v
(z'\,z)\rangle \eqno(II.24)
$$
where $\lambda$ is the coefficent to be defined later. Taking into account Eq.
(II.23), and
approximating $\langle\land^+_\mu \land_\nu\rangle \cong (-1) \langle\psi^+_\nu
(z') \psi_\mu
(z')\rangle$ we find:

$$
i~\partial_{\bar z} A_-~=~\lambda\pi \langle\psi^+_\nu (z) \psi_\mu (z)\rangle
\eqno(II.25)
$$
or in terms of spin components:

$$
i~\partial_{\bar z} A^i_-~=~\lambda\pi~\langle S^i (z)\rangle \eqno(II.26)
$$
As we mentioned, the ground state expectation value of spin operator is zero.
Any spin
excitation, however, produce the gauge potential $A^i_-$. For example, for spin
1/2 quasihole
in Halperin-Haldane state $\langle S^z\rangle = \delta (z-z_o$) will lead to a
gauge
potential of a point-line source:

$$
\eqalignno{i~A^z_- (z')~=~&-~{\lambda\over 2}~\int~{d^2z\over
z-z^\prime}~\langle\psi^+
(z)~\sigma^z~\psi (z)\rangle\cr
\noalign{\vskip 1pc}
&=~-~{\lambda\over 2}~{1\over z_o - z^\prime}&(II.27)\cr}
$$
The value of $\lambda$ is fixed by the requirement to be consistent with semion
statistics of
spin 1/2 excitations (neglecting the phase coming from charge sector) and gives
$\lambda$=1.
 From Eq. (II.24) it follows that in SQHE phase the displacement between zeros
of wave
function and positions of particles leads to a nonlocal effect, revealed by
effective gauge
potential. Assuming that scalar interactions in the system are short ranged, we
can write
down the local effective LG action whose variation leads to constraint Eq.
(II.24 $-$ II.26):

$$
\eqalignno{S~&=~\int d^2 x  dt~\land^+_\mu~(i~\partial_o 1^{\mu v} - A^{\mu
v})~\land_v~+~{1\over 4\pi}~Tr \hat A_i~\partial_j \hat A_k \epsilon^{ijk}\cr
\noalign{\vskip 1pc}
&+~V (\land^+\land)~+~{1\over 2M} \left|(i~\partial_i 1^{\mu v} - A^{\mu
v}_i)\land_v\right|^2&(II.28)\cr}
$$
Where we also take into account special gradients of the order parameter
$\land_\mu (z)$
defined in Eq. (II.19); potential $V (\land^+\land$) provides the fixed
amplitude of the
order parameter. The most nontrivial part of the effective LG action is the $Tr
\hat A_i
\partial_j \hat A_k \epsilon^{ijk}$ term, which is recognized as a gradient
part of the
SU(2)$_{k=1}$ Chern-Simons term:

$$
{\cal L_{CS}}~=~\int d^2 x dt~{k\over 4\pi}~Tr (\hat A_i~\partial_j~\hat
A_k~+~2/3~\hat
A_i~\hat A_j~\hat A_k) \epsilon^{ijk} \eqno(II.29)
$$
at k$\equiv$1 for our case (the subscript k in SU(2)$_k$ means precisely the
coefficient in
front of Chern-Simons term). The approximations we use does not allows us to
find the second
term in Chern-Simons Lagrangian Eq.(II.29). Locally this term always can be
gauged  out. However
it is important for global topological structure of the Chern-Simons term. It
is clear from
Eq.(II.26) that this term is a higher order correction in the gauge  we choose
deriving
Eq.(II.26).

It is reasonable to argue that because of local spin correlations in SQHE state
the true
SU(2) rotational invariance should be observed. Although above we identify the
spin
$\pm$ particles with flux $\pm \pi$ in state $\Psi^{(2)} ([z^+_i], [z^-_i])$
this
identification requires the spin quantization axis to be fixed explicitly. This
is the
``abelian'' way to incorporate spin quantum numbers of electrons into the wave
function.

 In this procedure the single particle states are described in
terms of the
spin projection on the z-axis, and for simplicity, this axis is assumed to be
in
the same direction everywhere. Thus, we deal only with the U(1)
diagonal subgroup of the full SU(2) spin group. Also, the plasma analogy for
$ \q$ leads to the correspondence  with the
two component plasma with effective charge $q =+ q_0$ for spin$\uparrow$ and $q
= -q_0$ for spin $\downarrow$ particles. This analogy suggests that we should
attach different fluxes to particles with opposite spin and deal with them
 in much the same way as we did with the charge sector in section II.

However, there is a problem with this approach.
So far there is no spin anisotropy in this state since we have  neglected
 the Zeeman term
in the consideration of the SQHE \refto{Hal},. The ``abelian" approach
breaks the SU(2) spin symmetry from the outset. Its recovery is a highly
non-trivial matter. In principle one
has to be able to formulate the SQHE wave function  while keeping the full
SU(2) invariance and to allow for a
quantization axis that is varying in space.
Girvin et al. \refto{G} have pointed out that  $\q$
leads to a partition function for a two-component plasma and that any
extra charge
= spin is screened. The screening in the two component anyon gas, in the
context of the spin coupled to a gauge field, was found in
reference \cite{BK}. Thus, what is needed is a procedure to attach different
fluxes to particles with $\uparrow$ and $\downarrow$ spins in a manner that is
compatible with the SU(2) spin symmetry. Fortunately such an approach does
exist: it is the non-abelian SU(2) CS theory. A non-abelian CS term, much like
the abelian CS theory used in the description of the spin
polarized
QHE \refto{GM,ZHK,R,LF}, attaches fluxes to particles. But, unlike the
``abelian" approach mentioned above, the non-abelian CS theory is invariant
under SU(2) rotations of the spin. Furthermore, this invariance is local
and the theory is a gauge theory. It turns out that the CS theory represents
the
only possible local way to attach particles to SU(2) fluxes.
Below we will follow this second way in considering the spin wave function.

Consider
the set of coordinates $[\zu], [\zd]$ of a set of some spinors with
the spin up components, located at points$[\zu]$, and spin down at points
$[\zd]$. The points $[\zu],[\zd]$ will be regarded as
the positions of sources of an SU(2) field $\land_\mu$, taken in the
fundamental
representation. It  corresponds to the spin 1/2 of the electrons , constituing
the QHE state. The
Lagrangian $\cL_{\rm spin}$ of the spin sector is  given by Eq.(II.28)
with the full  non-abelian Chern-Simons term.

The points at which the excitations  are located are the
the sources for the gauge field . As it can be seen from
the variation of the Lagrangian
(II.28) over $A_0^a$:
$$ {\delta \cL_{\rm spin} \over{\delta A_0^a}} = \land^+\sigma^a\land +
{k\over{\pi}}F^a_{xy} = 0.\eqno(II.30)$$
The strength of the gauge field is given by $F^a_{xy} = \partial_xA^a_y
- \partial_yA^a_x + [A_x , A_y]^a$.
Let us assume that the particles have a mass $m$. The path-integral
representation of a matrix element of the evolution operator is given as a sum
over all possible particle trajectories and gauge field histories. The
constraint of Eq.(II.30) requires that each term in this amplitude
should contain a factor representing a path-ordered exponential of the
$SU(2)$ gauge field along each particle trajectory. These path-ordered
exponentials are usually referred to as Wilson lines. In first quantization,
the
time evolution during the  time interval $t$ of the heavy sources will be given
 by the
amplitude:
$$\eqalign{ \Psi([z'^+_i],[z'^-_i], t) =
&\sum_{Paths} e^{-i\int dt (\sum_i m/2
|dz^+_i/dt|^2 + \sum_{i} m/2 |dz^-_i/dt|^2)} \cr \int D[A] &
\otimes_{i,j} W_i(z'^+_i,z^+_i) W_{j}(z'^-_j, z^-_j)
e^{ik\int d^2x dt \cL_{CS}} \Psi([\zu],[\zd],0) \cr } . \eqno(II.31)$$
where $z'^+ , z'^-$ are the set of final positions of the sources, and
$$ W_i(z'_i, z_i) = [Pe^{i\int_{z_i}^{z'_i} A_ldx^l} ]. \eqno(II.32)$$
are Wilson lines evaluated on the 3-dimensional paths from $z_i$ to $z'_i$. We
will consider the 2-D disc geometry pierced by the Wilson lines. The
coordinate space is $ D\times R$, where R is the time.
The integral in the exponent in $W_i(z'_i, z_i)$ is the quasiclassical
 expression
for the spin- current- gauge potential coupling $\int A^a_{\mu}j^a_{\mu} d^2x
dt$, assuming that $ j^a_{\mu} = \sigma^a {dx_{\mu}\over dt }  \delta (x -
x_l(t))$ and $x_l(t)$ parametrizes the quasiclassical path of the particle.

The CS action for the gauge field leads to the effective semion statistics of
Wilson lines. Let us fix two Wilson lines, corresponding, for example to
particles at $z^+_1$ and $z^-_1$. And let us consider two processes
which represent evolutions
with the same final state and only differ by the presence of an extra knot in
their histories given by  $W(z'^+_i, z^+_1), W(z'^-_1, z^-_1)$ . Then the
final amplitudes $
\Psi([z'^+_i],[z'^-_i])$ will gain different phases in these processes.
One can find \refto{Labastida}, that the amplitudes are related by
$$ \Psi_{knotted}([z'^+_i],[z'^-_i]) = \exp(i \gamma)
\Psi_{unknotted}([z'^+_i],[z'^-_i])\eqno(II.33)$$
where $\gamma$ is the {\it conformal weight} of the primary field for the
$SU(2)$ level $k$ group, and is given by
$$ \gamma = {4\pi j (j+1)\over{ k+ 2}} \eqno(II.34)$$
In our case, $k=1, j= \12$, the phase difference between two
configurations
is $\pi$ which corresponds to a phase of $\pi/2$ per particle. If we assume
that the
evolution between two configurations is adiabatic, the kinetic energy does
not modify the value of $\gamma$ because it is quadratic in time
derivative. The only contribution to the phase comes from the CS action and it
 leads to the semion statistics of the excitations
, exhibited in the spin wave function
$\q$ \refto{B}.

\head {III.Conclusion}
\vskip 0.5pc

In this paper the theory of SQHE phase was presented. We considered the
charge-spin
factorization in the Halperin-Haldane state and argue that the Halperin-Haldane
state
variational wave function can be written in the form of product of two wave
functions: one,
$\Psi^{(1)}_m ([z^+_i], [z^-_i])$ corresponds to charged spinless semions in
external
magnetic field and the other --- spin wave function $\Psi^{(2)} ([z^+_i],
[z^-_i])$ is the
wave function of spin 1/2 neutral semions. Because all states in the Hilbert
space of the
problem can be represented as a direct product of charge and spin contribution
we argue that
LG theory of SQHE phase can be written as the theory for composite order
parameter $\land_\pm
(z)$ from Eq. (II.20). In our derivation of the LG theory we concentrate on the
spin sector.
The parts of LG action for charge sector can be obtained, following the
derivation of LG
theory for spin polarized FQHE.

We construct the coherent states for SQHE phase which are analogous to the
coherent states
for polarized FQHE phase, and describe SQHE  as the phase with undefined spin
projection and
undefined number of particles. The SQHE order parameter has a nonzero diagonal
expectation value
in this coherent state.

Because of the ``spin rigidity'' of SQHE state, the nodes of spin wave function
$\Psi^{(2)}
([z^+_i], [z^-_i])$ are confined to the positions of the particles. Moreover we
find that any
displacements of these nodes from positions of the particles, described by
nonlocal composite
operator $\land_\mu (z',z) = \psi^+_\mu (z') \prod_{\mu^\prime}
U^{1/2\mu\cdot\mu^\prime}_{\mu^\prime} (z)$ leads to nontrivial monodromy of
the wave function
around closed contour, enclosing such a displacement. The natural measurement
of ths
monodromy of spin wave function is the SU(2) valued gauge potential $\hat A_i$
with the flux
$F^i_{xy}$ proportional to the noncompensated spin density $\langle\hat
S^i\rangle$. Although
we were not able to reproduce full SU(2) invariant topological term, we argue
that because of
local SU(2) invariance in SQHE phase, the spin part of LG action contains
SU(2)$_{k=1}$ Chern
Simons term. We also find that the topological structure of the Chern-Simons
theory leads
unambiguously to semion statistics of excitations in the spin sector of SQHE.
However, these
excitations are not physically relevant, because in the bulk of SQHE phase
spinons are
confined with holons in order to have trivial monodromy for Halperin-Haldane
wave function,
written in terms of electron coordinates. It has been argued in\refto{BS}, that
SU(2)$_{k=1}$
Kac-Moody algebra, closely related with the $SU(2)_{k=1}$ Chern-Simons action
describe the edge of SQHE. This current algebra  leads to the neutral spinon
excitations as
part of the Hilbert space of the edge.

This consideration of S=1/2 SQHE state is also useful in revealing the
connection between
conformal field theory and different phases of FQHE.\refto{MR,BF} For example
using these
results we can show that i) there is a S=1 Singlet QHE variational wave
function, ii) this
wave function is given by conformal block of SU(2)$_{k=2}$ Chern-Simons theory,
iii) it
supports nonabelian excitations with fractional charge and spin
1/2.\refto{unpublished}

\vskip 1pc

\noindent
ACKNOWLEDGMENTS
\vskip 0.5pc

We would like to thank E. Fradkin, F. D. M. Haldane, M. Stone and N. Read for
collaboration
and useful discussions. This work was supported in part by J. R. Oppenheimer
Fellowship and
Department of Energy.

\endpage

\references

\refis{BS} A. V. Balatsky and M. Stone, \prb 43, April, 1991.

\refis{unpublished} A. V. Balatsky, unpublished

\refis{5/2} R. Willett et. al., \prl 59, 1776, 1987;
J. P. Eisenstein et. al., \prl 61, 997,
1988.


\refis{Str.Cor} G.Baskaran, Z.Zou and P.W.Anderson, \journal Solid State
Comm., 63, 973, 1987;
 S.Kivelson, D.Rokhsar and J.Sethna, \prb 35, 8865, 1987.

\refis{G} S. M. Girvin, A. H. Mac-Donald, M. P. A. Fisher, S.-J. Rey and J.
Sethna, \prl 65, 1671, 1990.

\refis{BK} A.V.Balatsky and V.Kalmeyer , \prb 43, 6228, 1991

\refis{Laughlin} R. B. Laughlin, \prl 50, 1395, 1983.

\refis{MR} G.Moore and N.Read, Yale University preprint (1990).

\refis{Hal} F.D.M.Haldane  Chapter 7 in ``The Quantum Hall Effect",
R.Prange and S.Girvin, Editors, Springer-Verlag (1990).



\refis{QHE} ``The Quantum Hall Effect", R.Prange and S.Girvin Editors,
 Springer-Verlag (1990).

\refis{GM} S.Girvin and A.MacDonald, \prl 58, 1252, 1987.

\refis{ZHK} S.C.Zhang, T.Hansson and S.Kivelson, \prl 62, 82, 1989.

\refis{Halperin} B.I.Halperin, \prl 52, 1583, 1984.








\refis{LF} A.Lopez and E.Fradkin, in preparation.




\refis{R} N. Read, \prl 62, 86, 1989.


\refis{Labastida} J. M. Labastida and F. Ramallo, CERN Preprint 1989.

\refis{B} A. V. Balatsky, \prb 43, 1257, 1991.

\refis{BF} A. V. Balatsky and E. Fradkin, \prb 43, 10622, 1991.

\refis{B2} A. V. Balatsky, \prl 66, 814, 1991.

\endreferences

\endit